\pdfoutput=1
\documentclass[11pt,twoside,a4paper,cmspaper,final,collab]{cms-tdr}
\def\svnVersion{8b006a2-D}\def\svnDate{2021/03/11}\def\cmsCernNoTag{CERN-EP-2020-173}\def\cmsCernDate{\today}\def\cmsMessage{"Published in the Journal of High Energy Physics as \href{http://dx.doi.org/10.1007/JHEP03(2021)011}{\doi{10.1007/JHEP03(2021)011}.}"}
\begin{document}\cmsNoteHeader{EXO-20-005}

\newcommand{\vll}{\ensuremath{\overrightarrow{\ell\ell}}\xspace}
\newcommand{\vmetg}{\ensuremath{\pt^{\ptvecmiss+\vg}}\xspace}
\newcommand{\mll}{\ensuremath{m_{\ell\ell}}\xspace}
\newcommand{\absetag}{\ensuremath{\abs{\eta^\gamma}}\xspace}
\newcommand{\wj}{\ensuremath{\PW+\text{jets}}\xspace}
\newcommand{\wenj}{\ensuremath{\PW(\Pe\PGn)+\text{jets}}\xspace}
\newcommand{\zj}{\ensuremath{\PZ+\text{jets}}\xspace}
\newcommand{\gj}{\ensuremath{\PGg+\text{jets}}\xspace}
\newcommand{\wlng}{\ensuremath{\PW(\ell\PGn)+\PGg}\xspace}
\newcommand{\zmmg}{\ensuremath{\PZ(\PGm^+\PGm^-)+\PGg}\xspace}
\newcommand{\wmng}{\ensuremath{\PW(\PGm\PGn)+\PGg}\xspace}
\newcommand{\vg}{\ensuremath{\ptvec^{\gamma}}\xspace}
\newcommand{\zg}{\ensuremath{\PZ+\PGg}}
\newcommand{\ptg}{\ensuremath{\pt^{\gamma}}\xspace}
\newcommand{\jet}{\ensuremath{\mathrm{j}}}
\newcommand{\detajj}{\ensuremath{\abs{\Delta\eta_{\jet\jet}}}\xspace}
\newcommand{\mjj}{\ensuremath{m_{\jet\jet}}\xspace}
\newcommand{\dphijetmet}{\ensuremath{\Delta \phi_{\text{jet},\ptvecmiss}}\xspace}
\newcommand{\pttot}{\ensuremath{\pt^{\text{tot}}}\xspace}
\newcommand{\sieie}{\ensuremath{\sigma_{\eta \eta}}\xspace}
\newcommand{\ttg}{\ensuremath{\ttbar\PGg}\xspace}
\newcommand{\tg}{\ensuremath{\PQt\PGg}\xspace}
\newcommand{\zinvg}{\ensuremath{\PZ(\Pgn\Pagn)+\PGg}\xspace}
\newcommand{\zep}{\ensuremath{z_{\gamma}^{*}\xspace}}
\newcommand{\sigmavbf}{\ensuremath{\sigma_{\text{VBF}}}\xspace}
\newcommand{\sigmaSM}{\ensuremath{\sigma_{\text{SM}}}\xspace}
\newcommand{\brhinvg}{\ensuremath{\mathcal{B}(\PH\to\text{inv.}+\gamma)}\xspace}
\providecommand{\bigabs}[1]{\ensuremath{\Bigl|#1\Bigr|}\xspace}
\providecommand{\cmsTable}[1]{\resizebox{\textwidth}{!}{#1}}
\newlength\cmsTabSkip\setlength{\cmsTabSkip}{1ex}

\cmsNoteHeader{EXO-20-005} 

\title{Search for dark photons in Higgs boson production via vector boson fusion in proton-proton collisions at $\sqrt{s}=13\TeV$}
 
\date{\today}

\abstract{
A search is presented for a Higgs boson that is produced via vector boson fusion and that 
decays to an undetected particle and an isolated photon. 
The search is performed by the CMS collaboration at the LHC, 
using a data set corresponding to an integrated luminosity of 130\fbinv, 
recorded at a center-of-mass energy of 13\TeV in 2016--2018. 
No significant excess of events above the expectation from the standard model 
background is found. The results are interpreted in the context of a theoretical 
model in which the undetected particle is a massless dark photon. 
An upper limit is set on the product of the cross section for production via vector boson fusion 
and the branching fraction for such a Higgs boson decay, 
as a function of the Higgs boson mass. 
For a Higgs boson mass of 125\GeV, assuming the standard model production rates, 
the observed (expected) 95\% confidence level upper limit on the branching fraction is 3.5 (2.8)\%.
This is the first search for such decays in the vector boson fusion channel. Combination with
a previous search for Higgs bosons produced in association with a Z boson results in an observed
(expected) upper limit on the branching fraction of 2.9 (2.1)\% at 95\% confidence level. 
}

\hypersetup{%
pdfauthor={CMS Collaboration},%
pdftitle={Search for dark photons in Higgs boson production via vector boson fusion in proton-proton collisions at sqrt(s) = 13 TeV},%
pdfsubject={CMS},%
pdfkeywords={CMS,  dark photons}}

\maketitle

\section{Introduction}
\label{sec:Introduction}

Following the observation of a Higgs boson by the ATLAS and CMS 
collaborations~\cite{AtlasPaperCombination, CMSPaperCombination, CMSPaperCombination2}, 
an important focus of the CERN LHC physics program has been the study of the properties of this particle.
The observation of a sizable branching fraction of the Higgs boson to invisible 
or almost invisible final states~\cite{Ghosh:2012ep,Martin:1999qf,Bai:2011wz,Gori} would be 
a strong sign of physics beyond the standard model (BSM). 
Studies of the new boson at a mass of about 125\GeV~\cite{Aad:2015zhl,Sirunyan:2017exp} show no significant 
deviation from the standard model (SM) Higgs boson hypothesis, and measurements of its couplings 
constrain its partial decay width to undetected decay modes~\cite{Khachatryan:2016vau,Sirunyan:2018koj}. 
Assuming that the couplings of the Higgs boson to $\PW$ and $\PZ$ bosons are not larger than
the SM values, an upper limit of 38\% has been obtained at 95\% confidence level ($\CL$) on the branching fraction of the 125\GeV Higgs boson to BSM 
particles by the CMS collaboration using data collected in 2016~\cite{Sirunyan:2018koj,Aad:2019mbh}.

\begin{figure}[htbp]
\centering
\includegraphics[width=0.42\textwidth]{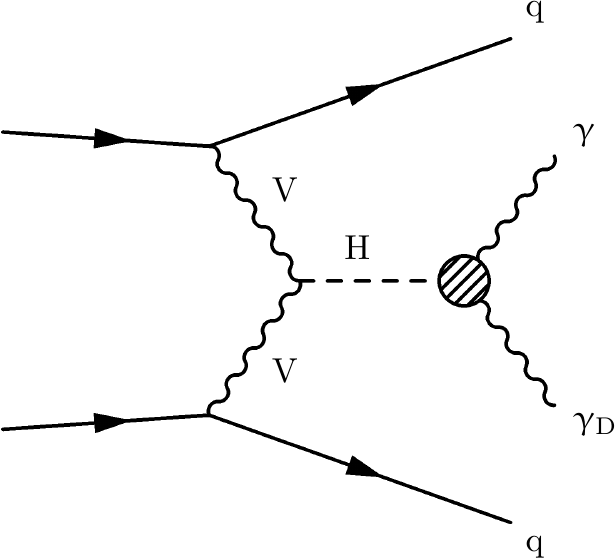}
\caption{A Feynman diagram for the VBF production of the 
$\Pq\Pq\PH(\gamma\gamma_\mathrm{D})$ final state.\label{fig:exo-higgs-decay}}
\end{figure}

This paper presents a search for a scalar Higgs boson $\PH$ produced via vector boson fusion (VBF) 
and decaying to an undetected particle and a photon $\PGg$. 
Such Higgs boson decays are predicted by several BSM models~\cite{Gori,Djouadi:1997gw,Petersson:2012dp}. 
In this search, the target channel is $\Pq\Pq\PH(\to \gamma \gamma_\mathrm{D})$, 
where the final-state quarks ($\Pq$) arise from the VBF process and $\gamma_\mathrm{D}$ is a massless dark photon that couples to the Higgs boson through a 
dark sector~\cite{Gabrielli:2013jka,Gabrielli:2014oya,Biswas:2016jsh,Biswas:2017anm}.
The dark photon escapes undetected. 
A Feynman diagram for this process is shown in Fig.~\ref{fig:exo-higgs-decay}. 
The branching fraction for a Higgs boson decaying to such an invisible particle and a photon, 
$\brhinvg$, could be as large as 5\% and still be consistent 
with current experimental constraints~\cite{Gabrielli:2014oya}. 
While the main focus of this search is on production via VBF, the additional contribution 
from gluon fusion production ($\Pg\Pg\PH$) is sizable if initial-state gluon radiation mimics the 
experimental signature of VBF. Thus, the $\Pg\Pg\PH$ process is also considered for the SM Higgs boson. Additionally, a model-independent search for VBF production 
is performed for heavy neutral Higgs bosons with masses between 125 and 1000\GeV~\cite{deFlorian:2016spz}, 
since similar decays are also possible for potential non-SM scalar bosons.

In the VBF production mode, a Higgs boson is accompanied by two jets that exhibit a large separation
in pseudorapidity ($\detajj$) and a large dijet mass ($\mjj$). This characteristic signature
allows for the suppression of SM backgrounds, making the VBF channel a very sensitive mode in the search for exotic Higgs boson decays. 
The invisible particle together with the photon produced in the Higgs boson decay can 
recoil with high transverse momentum (\pt) against the VBF dijet system, 
resulting in an event with a large missing transverse momentum (\ptmiss) which can be used to 
select signal-enriched samples.

The analysis summarized in this paper uses proton-proton ($\Pp\Pp$) collision 
data collected at $\sqrt{s}=13\TeV$ with the CMS detector in 2016--2018, with a 
total integrated luminosity of 130\fbinv. 
Similar searches have previously been performed by the CMS Collaboration using the data collected 
at $\sqrt{s}=8\TeV$~\cite{Khachatryan:2015vta} and  $\sqrt{s}=13\TeV$~\cite{Sirunyan:2019xst}, 
where the Higgs bosons were produced by $\Pg\Pg\PH$ or in association with a $\PZ$ boson, 
respectively. This analysis presents the first search for Higgs bosons decaying to an 
undetected particle and a photon using the VBF signature for Higgs boson production.

\section{The CMS detector}
\label{sec:cms}

The central feature of the CMS apparatus is a superconducting solenoid of 
6\unit{m} internal diameter, providing a magnetic field of 3.8\unit{T}. Within 
the solenoid volume are a silicon pixel and strip tracker, a lead tungstate 
crystal electromagnetic calorimeter (ECAL), and a brass and scintillator hadron 
calorimeter, each composed of a barrel and two endcap sections. 
The tracker system measures the momentum of charged particles in the region up to $\abs{\eta}<2.5$, 
where $\eta$ is the pseudorapidity, while the ECAL and HCAL provide coverage 
up to $\abs{\eta}<3.0$. Forward calorimeters extend the $\eta$ coverage provided by the barrel 
and endcap detectors to $\abs{\eta}<5.0$. Muons are detected in gas-ionization chambers embedded in 
the steel magnetic flux-return yoke outside the solenoid, which cover the region up to $\abs{\eta}<2.4$.

Events of interest are selected using a two-tiered trigger 
system~\cite{Khachatryan:2016bia}. The first level (L1), composed of custom hardware 
processors, uses information from the calorimeters and muon detectors to select 
events at a rate of around 100\unit{kHz} within a fixed time interval of less than 4\mus. 
The second level, known as the high-level trigger (HLT), consists of a farm of 
processors running a version of the full event reconstruction software optimized 
for fast processing and reduces the event rate to around 1\unit{kHz} before data storage.

A more detailed description of the CMS detector, together with a definition of the 
coordinate system used and the relevant kinematic variables, can be found in 
Ref.~\cite{Chatrchyan:2008aa}.

\section{Data samples and event reconstruction}
\label{sec:objects}

The data used in this search were collected in separate LHC operating periods 
in 2016--2018. 
The three data sets are analyzed independently, with calibration constants and correction factors 
appropriate for the LHC running conditions and CMS detector properties in each year.

Monte Carlo (MC) simulated events are used to model the expected signal and background 
yields. The dominant background processes are from $\wj$ and $\gj$ production, in addition 
to smaller contributions from $\wlng$, $\zg$, and $\zj$ processes. 
For each process, three sets of simulated events are 
needed to match the different data-taking conditions in each of the three years. 
The next-to-leading order (NLO) $\POWHEG$ v2~\cite{Frixione:2002ik,Nason:2004rx,Frixione:2007vw,Alioli:2008gx,Alioli:2010xd} 
generator is used to simulate the VBF and $\Pg\Pg\PH$ Higgs boson production processes at NLO in quantum chromodynamics (QCD), 
as well as the $\ttbar$, $\PQt\PW$, $\ttbar\gamma$, 
triple vector boson ($\PV\PV\PV$), and $\PW\PW$, $\PW\PZ$, and $\PZ\PZ$ ($\PV\PV$) processes. 
For the VBF signal process, the Higgs boson production cross section as a function of $m_{\PH}$, incorporating the inclusive next-to-NLO 
QCD and NLO electroweak corrections, is taken from Refs.~\cite{Heinemeyer:2013tqa,deFlorian:2016spz}, where an SM-like Higgs boson is assumed. Monte Carlo events 
with SM-like Higgs boson masses of $m_{\PH}=125$, 150, 200, 300, 500, 800, and 1000\GeV are simulated. 
The semi-visible decay of the Higgs boson $\PH \to \gamma \gamma_\mathrm{D}$ is
simulated with $\PYTHIA$ 8.226 (8.230) for the 2016 (2017--18) sample~\cite{Sjostrand:2014zea}. The same versions of $\PYTHIA$ are used 
to simulate the parton showering and hadronization for all processes.
The $\wj$, $\zj$, and $\gj$ background processes are generated using $\MGvATNLO$ 2.2.2 (2.4.2) at leading order (LO) accuracy in QCD with up to four partons for 2016 (2017--18) \cite{Alwall:2014hca}. The different jet multiplicities of these samples are merged using the MLM scheme~\cite{MLMmerging} 
to match matrix element and parton shower jets. 
The LO simulations for these processes are corrected using boson $\pt$-dependent NLO QCD 
K-factors derived using $\MGvATNLO$. They are also corrected using $\pt$-dependent higher-order electroweak 
corrections extracted from theoretical calculations~\cite{Lindert:2017olm}. 
Production of $\wlng$ and $\zg$ events with up to one additional parton is simulated at 
NLO accuracy in QCD using the $\MGvATNLO$~2.2.2 (2.4.2) generator with the FxFx scheme~\cite{Frederix2012} for 2016 (2017 and 2018) samples. 
The same generator without the FxFx scheme is used to model the electroweak production of $\wlng$, $\wj$, $\zg$, 
and $\zj$ events with two partons at LO precision in QCD.
The NNPDF 3.0 NLO \cite{Ball:2014uwa} (NNPDF 3.1 next-to-next-to-leading order~\cite{Ball:2017nwa}) 
parton distribution functions (PDFs) are used for simulating all 2016 (2017--18) samples. 
The modeling of the underlying event is generated using the 
CUETP8M1~\cite{Skands:2014pea,Khachatryan:2015pea} and CP5 tunes~\cite{Sirunyan:2019dfx} 
for simulated samples corresponding to the 2016 and 2017--18 data sets, respectively.

All MC generated events are processed through a simulation of the CMS detector based on 
\GEANTfour~\cite{Geant} and are reconstructed with the same algorithms used for data. 
Additional $\Pp\Pp$ interactions in the same and nearby bunch 
crossings, referred to as pileup, are also simulated. 
The distribution of the number of pileup interactions in the simulation is adjusted 
to match the one observed in the data. 
The average number of pileup interactions was 23 (32) in 2016 (2017--18).

The CMS particle-flow (PF) algorithm~\cite{Sirunyan:2017ulk} is used to combine 
the information from all subdetectors for particle reconstruction and identification.
Jets are reconstructed by clustering PF candidates using the anti-\kt 
algorithm~\cite{Cacciari:2008gp} with a distance parameter of 0.4. 
Jets are calibrated in the simulation, and separately in data, accounting for energy 
deposits of neutral particles from pileup and any nonlinear detector response~\cite{Khachatryan:2016kdb,CMS-DP-2020-019}. 
Jets with $\pt>30\GeV$ and $\abs{\eta}<4.7$ are considered in the analysis. 
The effect of pileup is mitigated through a charged-hadron subtraction technique, 
which removes the energy of charged hadrons not originating from the primary 
interaction vertex (PV)~\cite{Sirunyan:2020foa}. 
The PV is defined as the vertex with the largest value of summed 
physics-object $\pt^2$. Here, the physics objects are the jets clustered using the jet finding 
algorithm~\cite{Cacciari:2008gp,Cacciari:2011ma} with the tracks assigned to the candidate 
vertex as inputs, and the associated \ptvecmiss is 
calculated as the negative vector \pt sum of those jets.

For further analysis the vector \ptvecmiss is defined as the negative vector \pt sum of all PF 
particle candidates and its magnitude is defined as \ptmiss. 
Corrections to jet energies due to detector 
response are propagated to \ptvecmiss~\cite{Khachatryan:2014gga}. Events with 
possible contributions from beam halo processes or anomalous signals in the calorimeters are 
rejected using dedicated filters~\cite{Khachatryan:2014gga}. 

Electrons and muons are reconstructed by associating a track reconstructed 
in the tracking detectors with either a cluster of energy in the 
ECAL~\cite{Khachatryan:2015hwa,CMS-DP-2018-017} or a track in the muon system~\cite{Sirunyan:2018fpa}. 
Events are rejected from the signal region (SR) if any electron (muon) with $\pt>10\GeV$ and $\abs{\eta}<2.5$ (2.4) 
passing the ``loose'' identification criteria is found~\cite{Khachatryan:2015hwa,Sirunyan:2018fpa}. 
Several leptonic control regions are defined, where muons must pass the ``medium" 
identification and ``tight" isolation working points~\cite{Khachatryan:2015hwa}, while electrons must pass the 
``tight" identification and isolation working points~\cite{Sirunyan:2018fpa}. 
Section~\ref{sec:backgrounds} provides 
more details about the control regions used in the analysis.

Finally, photon candidates are reconstructed from energy deposits in the 
ECAL~\cite{Khachatryan:2015iwa} with $\abs{\eta}<1.47$ (barrel region) and $\pt>80\GeV$. 
The identification of the candidates is based 
on shower shape and isolation variables, and the medium working point, 
as described in Ref.~\cite{Khachatryan:2015iwa}, is chosen to select those candidates. 
For a photon candidate to be considered as isolated, scalar sums of the
transverse momenta of PF charged hadrons, neutral hadrons, and photons within a cone of 
$\Delta R = \sqrt{\smash[b]{(\Delta\eta)^2+(\Delta\phi)^2}}<0.3$ around 
the candidate photon must fall below certain bounds~\cite{Khachatryan:2015iwa}. Only the PF candidates 
that do not overlap with the candidate photon are included in the isolation sums. 
In addition, a standard ``pixel-seed electron veto"~\cite{Khachatryan:2015iwa} is applied to 
reject electrons misidentified as photons. 
The electron to photon misidentification rate is measured in $\PZ \to \Pe\Pe$ events 
by comparing the ratio of $\Pe\gamma$ to $\Pe\Pe$ pairs consistent with 
the $\PZ$ boson mass. The average misidentification rate is 2--3\%. 
If a jet overlaps with a reconstructed photon fulfilling loose identification criteria~\cite{Khachatryan:2015iwa}, 
the jet is removed.

\section{Event selection}
\label{sec:selection}

Collision events were collected using a dedicated VBF+$\gamma$ trigger in 2016, 
while in 2017--18, a combination of single-photon and \ptmiss triggers 
was used. The HLT algorithm in 2016 is seeded by an $\Pe/\gamma$ L1 object with 
a \pt threshold of 40\GeV and comprises two parts. In the first part, a photon 
is reconstructed in the barrel region around the L1 object, imposing initial requirements on shower shapes and isolation. 
The photon \pt must be greater than 75\GeV. In the second part, calorimeter towers in the event
are clustered into anti-\kt jets~\cite{Cacciari:2008gp} with a distance parameter of 0.4. The event is 
recorded if it contains a pair of jets with $\pt>50\GeV$, with $\mjj>500\GeV$ and $\detajj>3$. 
This trigger was available for most of the data recorded throughout 2016, and provided an effective integrated luminosity of 28.5$\fbinv$. 
This dedicated trigger made possible the offline selection of events with much lower photon $\pt$ and \ptmiss than could be achieved 
with the single-photon and \ptmiss triggers used in 2017 and 2018~\cite{Sirunyan:2018dsf,Sirunyan:2017jix}. These triggers 
required a photon at the HLT with $\pt>200\GeV$ and $\abs{\eta}<1.47$, or $\ptmiss>120\GeV$, respectively.
The single-photon trigger path is used if an event satisfies both triggers and a photon with $\pt>230\GeV$ is 
identified in the offline analysis. If no such photon is identified, the event may be selected by the \ptmiss trigger path.

The signal topology consists of two forward high-\pt jets consistent with 
VBF production, large \ptmiss, and an isolated high-\pt photon. 
The signal cross section is several orders of magnitude lower than that of the major reducible 
background processes, and therefore a stringent selection is required to obtain a sample of 
sufficient purity to define the SR. To be consistent with the expected topology, 
the selection requires leading and subleading jets with $\pt>50\GeV$, and 
at least one photon in the barrel region with $\ptg>80\,(230)\GeV$ for the VBF+$\gamma$ and \ptmiss (single-photon) trigger paths. 
In addition, events are required to have between two and five jets in total, where each jet has $\pt>30\GeV$ and $\abs{\eta}<4.7$. 
For the purpose of rejecting the bulk of the $\gamma$+jets background, 
as well as the signal process with small Lorentz boost of the Higgs boson, 
a \ptmiss greater than 100 (140)\GeV in 2016 (2017--18) is required, and 
the azimuthal angle between all jets with $\pt>30\GeV$ and $\ptvecmiss$ 
($\dphijetmet$) must be $>$1.0. 
To reduce the background from leptonic events, a veto is applied rejecting events with 
any loosely identified electron or muon, as described in Section~\ref{sec:objects}.

To select the VBF topology, the two leading jets must be in opposite hemispheres, with $\detajj>3$ and $\mjj>500\GeV$, and the so-called
Zeppenfeld ($\zep$) variable~\cite{Rainwater:1996ud} must be $<$0.6, where
\begin{equation}
\zep \equiv \bigabs{\left(\eta_{\gamma}-(\eta_{\jet_1}+\eta_{\jet_2})/2\right)/\detajj},
\label{eq:zep}
\end{equation}
where $\eta_{\gamma}$ is the pseudorapidity of the photon, and $\eta_{\jet_{1}}$ and $\eta_{\jet_{2}}$ 
are the pseudorapidities of the two candidate VBF jets.  
Since the total \pt in the event should be consistent with zero, the modulus of the vector sum ($\pttot$) of
the \pt of the two leading jets, the \pt of the photon, and \ptmiss is required to be $<$150\GeV 
to reject events with jet \pt mismeasurement or with additional hard jets.
A summary of the SR selection for the analysis is shown in Table~\ref{tab:selectioncuts}. The different 
\ptmiss requirements on the three data sets are due to different data-taking conditions. 

\begin{table}[htbp]
  \centering
  \topcaption{Summary of the selection criteria in the SR, depending on the trigger path and data-taking year. 
  Rows with a single entry indicate that the same requirement is applied for all data-taking years and trigger paths.}
  \label{tab:selectioncuts}
  \begin{tabular} {lccc}
\hline
Data-taking year                   & 2016          & \multicolumn{2}{c}{2017/2018}  \\
\hline
Trigger                            & VBF+$\gamma$  & Single-photon & \ptmiss      \\
Number of photons                  & \multicolumn{3}{c}{$\geq$1 photon}     \\
$\ptg$                             & $>$80\GeV & $>$230\GeV & $>$80\GeV     \\
Number of leptons                  & \multicolumn{3}{c}{0}		    \\
$\pt^{\jet_1}$, $\pt^{\jet_2}$     & \multicolumn{3}{c}{$>$50\GeV}          \\
\ptmiss                            & $>$100\GeV & $>$140\GeV & $>$140\GeV   \\
Jet counting                       & \multicolumn{3}{c}{2--5}		    \\
$\mjj$                             & \multicolumn{3}{c}{$>$500\GeV}	    \\
$\detajj$                          & \multicolumn{3}{c}{$>$3.0} 	    \\
$\eta_{\jet_1}\eta_{\jet_2}$       & \multicolumn{3}{c}{$<$0}		    \\
$\dphijetmet$                      & \multicolumn{3}{c}{$>$1.0 radians}     \\
$\zep$                             & \multicolumn{3}{c}{$<$0.6} 	    \\
$\pttot$                           & \multicolumn{3}{c}{$<$150\GeV}	    \\
  \hline
  \end{tabular}
\end{table}

\section{Background estimation}
\label{sec:backgrounds}

There are multiple sources of SM background to the analysis. The most 
significant background arises from $\wenj$ production, where the photon candidate 
is a misidentified electron. For larger values of \ptmiss, the most important processes 
are the production of a photon with a $\PZ$ boson, where the $\PZ$ boson decays into a 
neutrino-antineutrino pair (\zinvg), and the production of a photon with a $\PW$ boson, where the $\PW$ boson decays to a lepton-neutrino pair (\wlng). 
For these processes, a VBF-like two-jet signature can be produced by initial-state QCD radiation.
The $\wlng$ process becomes an irreducible background if the charged lepton falls outside of the detector acceptance. 
Another significant background process is $\gj$ production with a mismeasured \ptmiss. 
Less significant background processes are $\PZ(\Pgn\Pagn)+\text{jets}$ and QCD multijet 
production, which can contribute to the SR when a jet is misidentified as a photon. 
For the $\wenj$, $\wlng$, \zinvg, and $\PZ(\Pgn\Pagn)+\text{jets}$ backgrounds, 
production via purely electroweak interactions, which is also considered, becomes more relevant at higher $\mjj$.

The main background processes described above are normalized by comparing the predicted yields 
to data in several control regions (CRs) defined to be as close as possible to the SR~\cite{Sirunyan:2017jix}. 
These regions are considered in the final discriminant maximum-likelihood fit, 
as described in Section~\ref{sec:fitting}. In particular, four CRs are defined:

\begin{itemize}
\item $\wenj$ region: the full SR selection is applied, except that an electron must 
be selected and no photons found, and the electron is then used in place of the 
signal photon to build all kinematic variables.

\item $\zmmg$ region: the full SR selection is applied, except that two muons must be 
selected together with a photon, and the $\dphijetmet$ requirement is not considered. 
The muons are added to $\ptvecmiss$ to emulate the signal topology.

\item $\wmng$ region: the full SR selection is applied, but a muon must be 
selected together with a photon, and the muon is added to $\ptvecmiss$ to 
emulate the signal topology. 

\item $\gj$ region: the full SR selection is applied, but $\dphijetmet$ must 
be $<$0.5.
\end{itemize}

There are other rare SM processes involving a photon and neutrinos or out-of-acceptance leptons, 
$\eg$ $\PV\PV$, $\PV\PV\PV$, $\ttg$, $\tg$. The contributions from these minor background processes are 
very small after the final selection, so they are estimated directly from MC simulation.

We also consider the possibility that a pathological event reconstruction could lead to a 
significant underestimation of the photon energy (mismeasured $\gamma$), leaving 
an event with large $\ptvecmiss$ aligned in azimuthal angle with a photon. 
These events can be selected as part of the SR and a yield estimation is needed. 
It is possible to model the distribution of such events using
the \gj\ simulation. Distributions obtained this way can be used in the signal extraction fit, 
as described below. Since the shapes of the kinematic distributions are sufficiently distinct between this background and
the signal, their rates can be determined simultaneously through the fit.

The distribution of this background is obtained by selecting events from \gj\ simulation with
the signal candidate selection criteria of Section~\ref{sec:selection}, 
excluding \ptmiss-related requirements. The content of these events is modified by setting 
the photon transverse momentum to a fraction of its original value 
and adding the difference in $\ptg$ to the $\ptvecmiss$ variable.
The nominal value for the new $\ptg$ used to obtain this background template is 50\%, and 
alternative scenarios using 25 and 75\% are considered to account for 
potential variations in the template shape. The overall normalization is assumed to have 
an uncertainty corresponding to a factor of two.

The rate of hadrons being misidentified as photons (nonprompt) is estimated using two low-\ptmiss $\gj$ 
samples~\cite{Sirunyan:2018dsf}. In the first sample, a binned template fit is performed 
on the distribution of the lateral extension of the ECAL shower of the photon candidate along the $\eta$
direction~\cite{Khachatryan:2015hwa}, $\sieie$, applying 
the full photon selection, except for the $\sieie$ requirement. Two sets of templates 
are created: for real photons and misidentified hadrons. The photon template is 
obtained using $\gj$ simulated events. The $\sieie$ distribution for the hadron template is derived 
from data using a sideband in the charged-hadron isolation distribution. 
The number of misidentified hadrons surviving the $\sieie$ requirement applied to 
the full photon selection is determined from the template fit. 
Their relative contribution to the total event yield in this low-\ptmiss sample is referred to as the hadron fake rate in the following.
The second low-\ptmiss sample, obtained by inverting the charged-hadron isolation 
requirement altogether and loosening the $\sieie$ requirement, 
almost exclusively consists of events with misidentified hadrons.
A hadron misidentification transfer factor is calculated as the ratio of the hadron fake rate
in the first subsample and the total yield in the second subsample. It is derived as a function of $\ptg$. 
The resulting misidentification transfer factors are then used to extrapolate to the SR from a high-\ptmiss 
control sample with the same photon candidate selection as applied for the second low-\ptmiss sample. 
An absolute prediction for the nonprompt background is then obtained by multiplying
the event yields in the control sample with the transfer factors. 
An uncertainty of 5 to 15\%, depending on the photon \pt, is assigned on the nonprompt rates to account for 
the limited statistical precision of the measurements. 
An alternative estimate of this background was made by considering events with $\mjj>500\GeV$.
An additional systematic uncertainty was assigned based on the observed difference 
between the two estimates.

\section{Signal extraction}
\label{sec:fitting}

After applying the selection, a binned maximum-likelihood fit to the transverse mass of the $\ptvecmiss$ and photon
system, $\mT$, is performed to discriminate between the signal and the remaining background processes, where $\mT$ is defined as
\begin{equation}
\mT \equiv \sqrt{\smash[b]{2\ptmiss\ptg[1-\cos(\Delta\phi_{\ptvecmiss,\vg})]}}, 
\label{eq:mt}
\end{equation}
and $\Delta\phi_{\ptvecmiss,\vg}$ is the azimuthal angle 
between the $\ptvecmiss$ and $\vg$ vectors. 
A profile likelihood technique is used where systematic uncertainties are represented by nuisance parameters~\cite{Cowan:2010js}. 
For each individual bin, a Poisson likelihood
term is used to describe the fluctuation of the yields around the expected
central value, which is given by the sum of the contributions from signal 
and background processes. 
The uncertainties affect the overall normalizations of the signal and background yields, 
as well as the shapes of the predictions across the distributions of the observables.
Uncertainties that affect only the normalization within a category are 
incorporated as nuisance parameters with log-normal probability density functions. 
Uncertainties affecting the template shapes are treated as nuisance parameters with Gaussian constraints. 
The normalization of each bin is interpolated smoothly with a sixth-order polynomial between the $\pm$1 
standard deviation variations and extrapolated linearly beyond this. 
The total likelihood is defined as the product of the 
likelihoods of the individual bins and the probability density functions 
for the nuisance parameters, including the product of the likelihood for the individual years. 

In addition, events in the SR and in all the CRs are split in two $\mjj$ regions, below and above 1500\GeV. 
This value is chosen to ensure roughly half of the VBF signal events are in each region. 
The division also makes it possible to account for different relative contributions to the 
$\wenj$, $\wlng$, \zinvg, and $\PZ(\Pgn\Pagn)+\text{jets}$ templates from strong or 
purely electroweak production mechanisms as a function of $\mjj$. 
In the $\zmmg$ and $\wmng$ CRs the $\mT$ variable emulates the one in the SR by adding the leptons to the \ptmiss. 
The exact $\mT$ binning choice in the SRs and CRs is summarized in Table~\ref{tab:binningdef}. 
Correlations between systematic uncertainties in different regions of $\mT$ and $\mjj$ used in the template fit 
are taken into account.
For all major background sources, normalization factors are used that are allowed to float freely in the fit. 
A single normalization factor for each process is used for the $\wlng$ and $\zg$ backgrounds, 
while for the $\wj$ and $\gj$ background processes, separate parameters are applied for each 
kinematic region defined by one bin of the respective CRs, resulting in six (two) 
separate normalization parameters for the $\wj$ ($\gj$) process. 
The events with mismeasured photons are included in the SRs as described in Section~\ref{sec:backgrounds}.

\begin{table}[htbp]
  \centering
  \topcaption{Summary of the $\mT$ binning choice in the SRs and CRs.}
  \label{tab:binningdef}
  \begin{tabular} {lcc}
\hline
 Region & Bins & $\mT$ range ({\GeVns}) \\
  \hline
SR, $\mjj<1500\GeV$             & 6 & [0, 30, 60, 90, 170, 250, $\infty$] \\
SR, $\mjj\geq1500\GeV$          & 6 & [0, 30, 60, 90, 170, 250, $\infty$] \\[\cmsTabSkip]

$\wenj$ CR, $\mjj<1500\GeV$	& 3 & [0, 90, 250, $\infty$] \\
$\wenj$ CR, $\mjj\geq1500\GeV$  & 3 & [0, 90, 250, $\infty$] \\
$\zmmg$ CR, $\mjj<1500\GeV$	& 1 & [0, $\infty$] \\
$\zmmg$ CR, $\mjj\geq1500\GeV$  & 1 & [0, $\infty$] \\
$\wmng$ CR, $\mjj<1500\GeV$	& 1 & [0, $\infty$] \\
$\wmng$ CR, $\mjj\geq1500\GeV$  & 1 & [0, $\infty$] \\
$\gj$ CR,   $\mjj<1500\GeV$     & 1 & [0, $\infty$] \\
$\gj$ CR,   $\mjj\geq1500\GeV$  & 1 & [0, $\infty$] \\
  \hline
  \end{tabular}
\end{table}

\section{Efficiencies and systematic uncertainties}
\label{sec:systematics}

Several sources of systematic uncertainty are taken into account in the 
maximum-likelihood fit. For each source of uncertainty,  the effects on the 
signal and background distributions are considered correlated.

The integrated luminosities of the 2016, 2017, and 2018 data-taking periods are 
individually known with uncertainties in the 2.3--2.5\% 
range~\cite{CMS-PAS-LUM-17-001,CMS-PAS-LUM-17-004,CMS-PAS-LUM-18-002}, while 
the total Run 2 (2016--2018) integrated luminosity has an uncertainty of 1.8\%.
The better precision of the overall luminosity measurement results from an 
improved understanding of relevant systematic effects.

The simulation of pileup events assumes a total inelastic $\Pp\Pp$ cross section of 69.2\unit{mb}, 
with an associated uncertainty of 4.6\%~\cite{ATLAS:2016pu,Sirunyan:2018nqx}, which has 
an impact on the expected signal and background yields of about 1\%.

Discrepancies in the lepton and photon reconstruction and identification
efficiencies between data and simulation are corrected by applying 
scale factors to all simulated samples.
These scale factors are determined using $\PZ \to \ell\bar{\ell}$ 
events in the $\PZ$ boson peak region that were recorded with 
unbiased triggers~\cite{Sirunyan:2018fpa,Khachatryan:2015hwa}. 
The scale factors depend on the \pt and $\eta$ of the lepton and have an uncertainty of 
${\approx}2$\% for both electrons and muons. 
The above procedure is applied also to determine the scale factors for photons using $\PZ \to \Pe^+\Pe^-$ events as a proxy,
and the yield uncertainty for photon candidates is ${\approx}4$\%. 
The photon momentum scale uncertainty is about 0.5\%. 
These uncertainties are treated as correlated across the three years.

The determination of the trigger efficiency leads to an 
uncertainty of ${\approx}1$\% in the VBF+$\gamma$ and single-photon triggers, 
while the uncertainty is ${\approx}7$\% for the \ptmiss triggers. 
These uncertainties are treated as uncorrelated across the 
three data sets and trigger selections, as data-taking conditions have varied across the three years.

The uncertainty in the calibration of the jet energy scale directly affects 
the acceptance of the jet multiplicity requirement and the \ptmiss measurement. 
These effects are estimated by shifting the jet energy in the simulation up and down by one standard deviation. 
The uncertainty in the jet energy scale is 2--5\%, depending on \pt and $\eta$~\cite{Khachatryan:2016kdb}, 
and the impact on the expected signal and background yields is about 3\%. 
The uncertainties in the jet energy scale are treated as uncorrelated across 
the three data sets.

The theoretical uncertainties due to the choice of  
QCD renormalization and factorization scales used in the simulation of the background processes are estimated 
by varying these scales independently up and down by a factor of 
two (excluding the two extreme variations) and taking the envelope of the resulting distributions 
as the uncertainty~\cite{Catani:2003zt,Cacciari:2003fi}. 
The variations of the PDF set and the strong coupling constant are used to estimate 
the corresponding uncertainties in the yields of the signal and background processes, 
following Refs.~\cite{Ball:2014uwa,Butterworth:2015oua}. 
The uncertainties in the signal predictions due to the choice of the
PDF set and the renormalization and factorization scale variations are taken from
Ref.~\cite{deFlorian:2016spz}. For the $\Pg\Pg\PH$ contribution, an
additional uncertainty of 40\% is assigned to take into account the
limited knowledge of the $\Pg\Pg\PH$ cross section in association with
two or more jets, as well as the uncertainty in the prediction of the 
$\Pg\Pg\PH$ differential cross section for large Higgs boson $\pt$, following the recipe
described in Refs.~\cite{Catani:2003zt,Sirunyan:2018owy}. 
Theoretical uncertainties in modeling the parton shower and underlying event primarily 
affect the jet multiplicity and are evaluated following the recipes from Refs.~\cite{Heinemeyer:2013tqa,deFlorian:2016spz}.

The statistical uncertainty associated with the limited number of simulated events 
is also considered a part of the systematic uncertainty. 
A summary of the impacts of the systematic uncertainties on the signal cross section for $m_{\PH}=125\GeV$ is presented in 
Table~\ref{tab:systematics}. The impacts are evaluated by fitting to Asimov data sets~\cite{Cowan:2010js} and are 
defined as the change in the fitted signal cross section when varying a nuisance parameter by its post-fit uncertainty. 
By performing the fit to the data simultaneously in the different CRs and SRs, 
the resultant final background uncertainties are reduced compared to the input uncertainties~\cite{Sirunyan:2017jix,Sirunyan:2018owy}. 
The impacts are shown for the case of a signal ($\sigma=0.05\sigmaSM$, where $\sigmaSM$ is the SM Higgs boson cross section
for $m_{\PH}=125\GeV$) and for the case of no signal ($\sigma=0$). 
The systematic uncertainties are dominated by the limited number of simulated events, 
the background normalization factors, and the jet energy scale.

\begin{table*}[htbp]
\centering
\topcaption{
Summary of the uncertainties in the fitted signal cross section in fb for $m_{\PH}=125\GeV$ 
assuming the presence of a signal ($\sigma=0.05\sigmaSM$) and the absence of a signal ($\sigma=0$).\label{tab:systematics}} 
{
\begin{tabular}{lcc}
\hline
\multirow{2}{*}{Source of uncertainty}       & Impact for scenario & Impact for scenario \\
                                             & with signal (fb)    & without signal (fb) \\
\hline
Integrated luminosity                        & 3.3   & 0.6 \\
Lepton and trigger measurements              & 17    & 7.7 \\
Jet energy scale and resolution		     & 24    & 19  \\
Pileup  				     & 9.7   & 8.5 \\
Background normalization		     & 25    & 18  \\
Theory            			     & 6.0   & 3.0 \\
Simulation sample size                       & 36    & 36  \\ [\cmsTabSkip]
Total systematic uncertainty                 & 54    & 46  \\
Statistical uncertainty                      & 58    & 48  \\ [\cmsTabSkip]
Total uncertainty                            & 79    & 66  \\
\hline
\end{tabular}
}
\end{table*}

\section{Results}
\label{sec:results}

The numbers of observed and expected events after applying the full selection requirements 
are shown in Table~\ref{tab:postfityields}. 
Owing to the anticorrelation between the yields of several background processes, the uncertainty 
in the background sum in the different regions is smaller than the uncertainties 
in some of the individual contributions. 
For illustration purposes, the signal shown has $\brhinvg$ set to $0.05$ and assumes 
the SM production cross section, as this corresponds roughly to the expected 
sensitivity level of the analysis.

\begin{figure}[hbtp]
\centering
\includegraphics[width=0.49\textwidth]{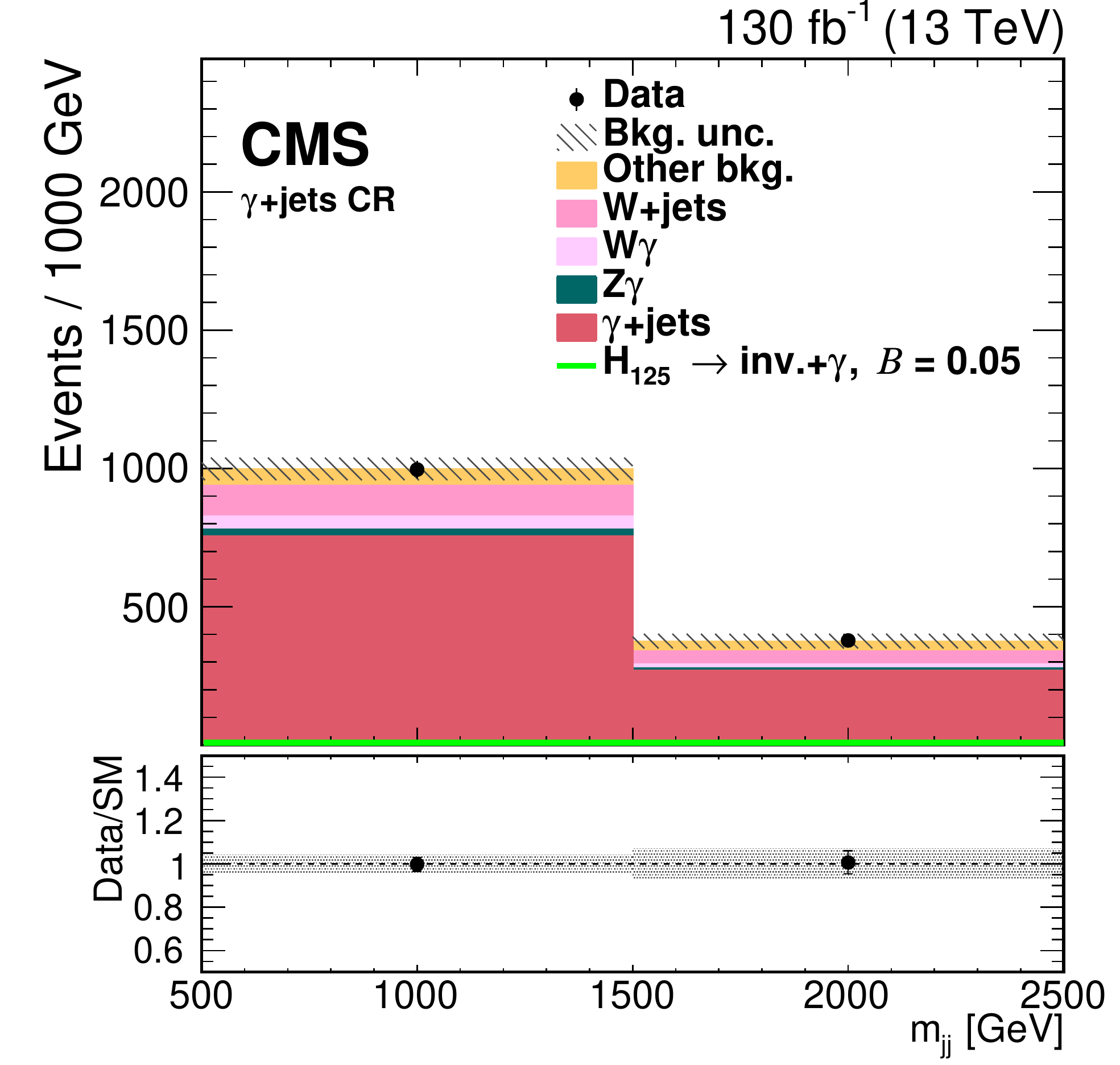}
\includegraphics[width=0.49\textwidth]{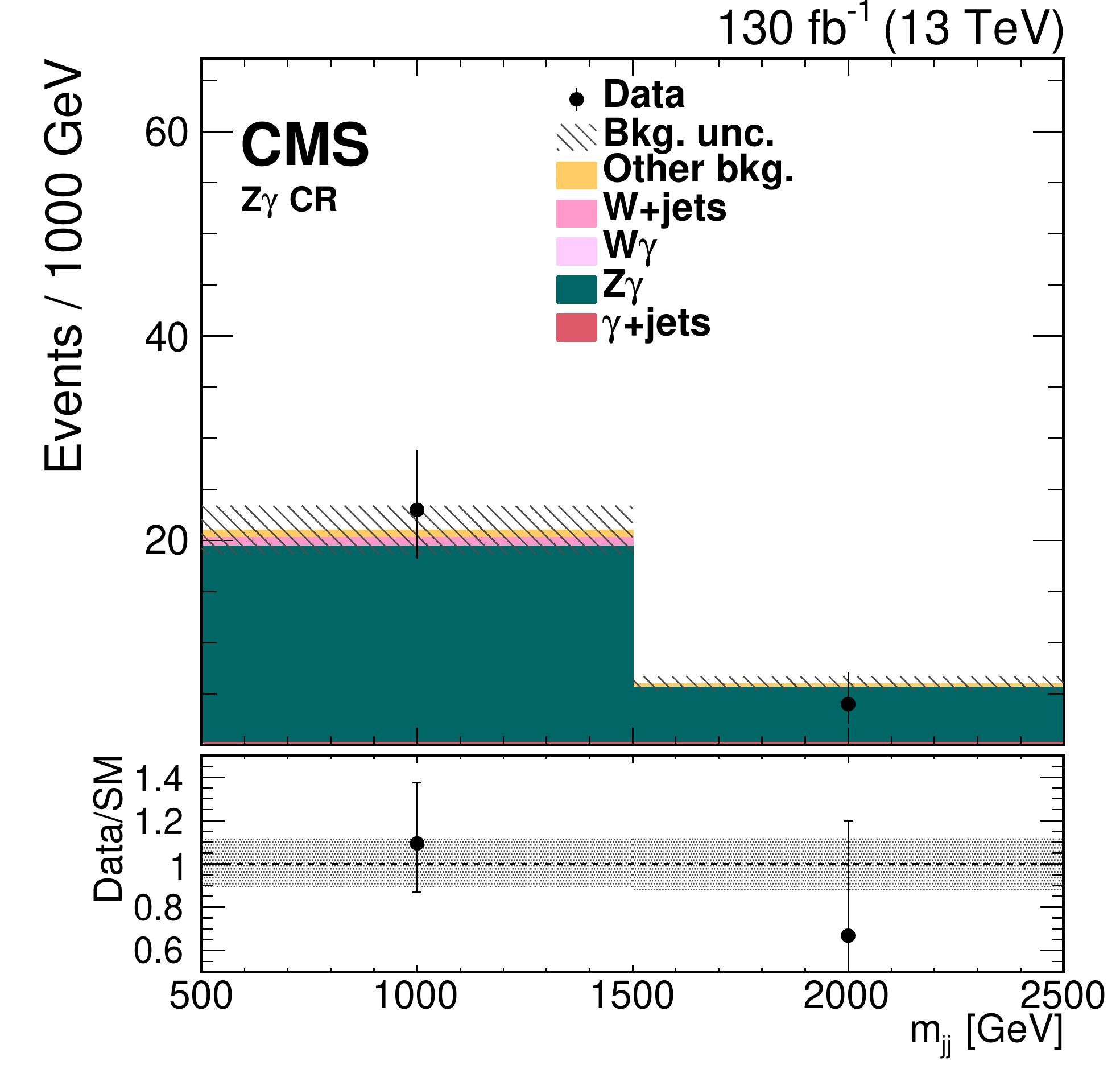}\\
\includegraphics[width=0.49\textwidth]{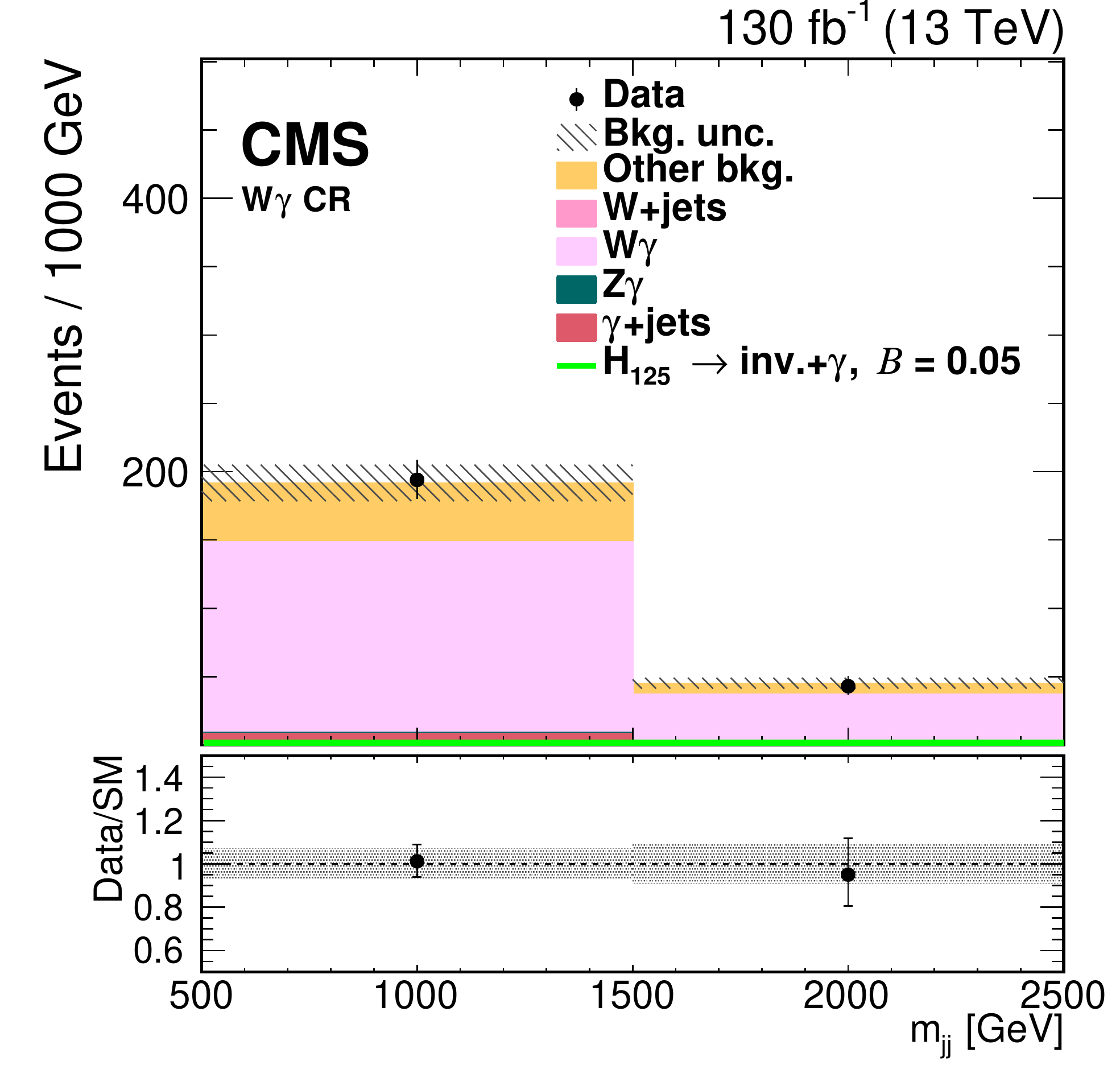}
 \caption{
The $\mjj$ distributions from the simultaneous fit in the 
$\gj$ (upper left), 
$\zmmg$ (upper right), 
and $\wmng$ (lower) CRs. The category other background 
includes contributions from $\zj$, nonprompt, top quark, $\PV\PV$, and $\PV\PV\PV$ processes. 
Overflow events are included in the last bin. The shaded bands 
represent the combination of the statistical and systematic uncertainties in the predicted yields. 
The light green line, illustrating the possible contribution expected from inclusive 
SM Higgs boson production, assumes a branching fraction of 5\% for $\PH\to\text{inv.}+\gamma$ decays.
The lower panel in the figures shows a per-bin ratio of the data yield and the background expectation.
The shaded band corresponds to the combined systematic and statistical uncertainty in the background expectation.
\label{fig:vbfg_comb0}}
\end{figure}

\begin{figure}[hbtp]
\centering
\includegraphics[width=0.49\textwidth]{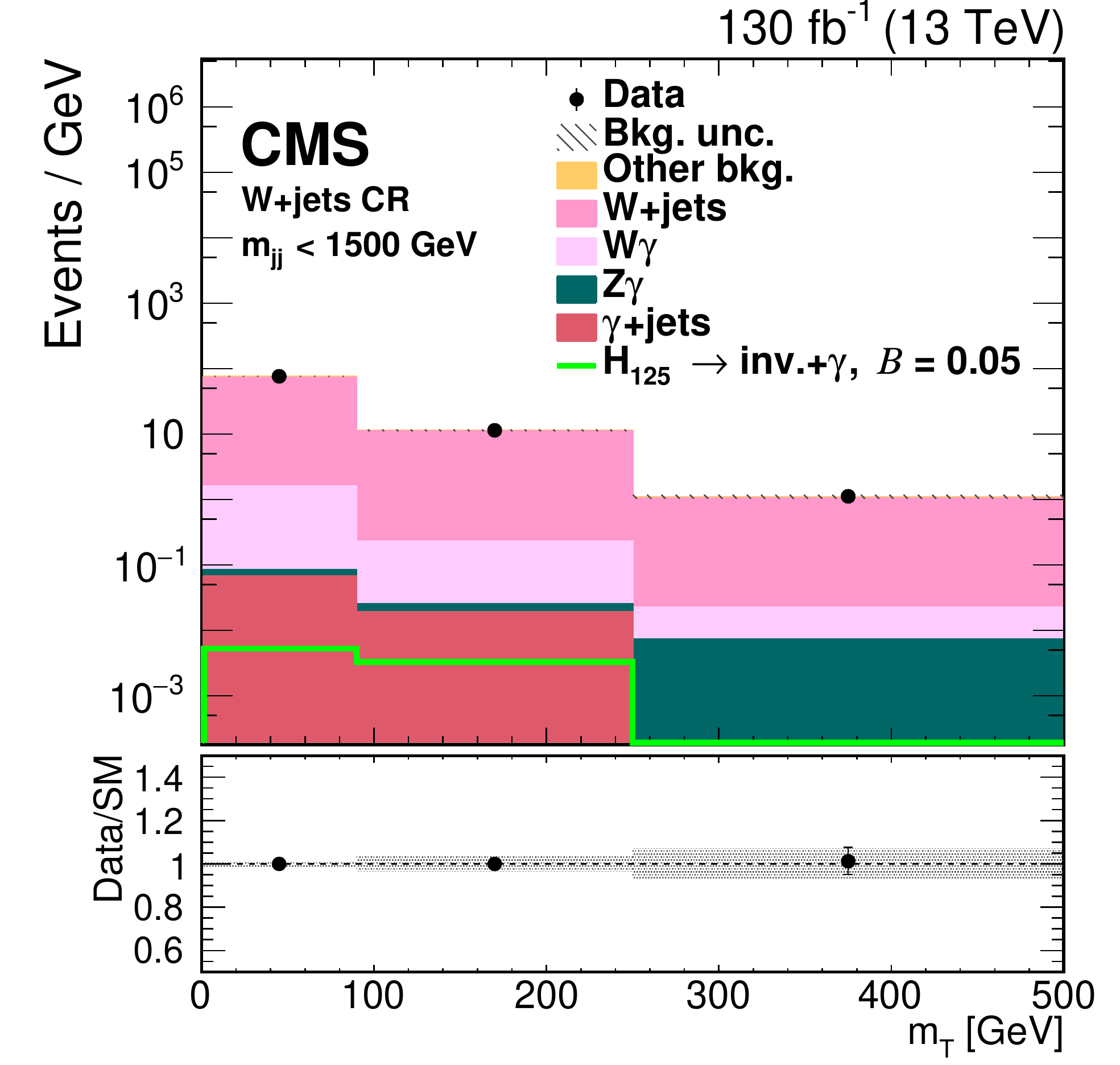}
\includegraphics[width=0.49\textwidth]{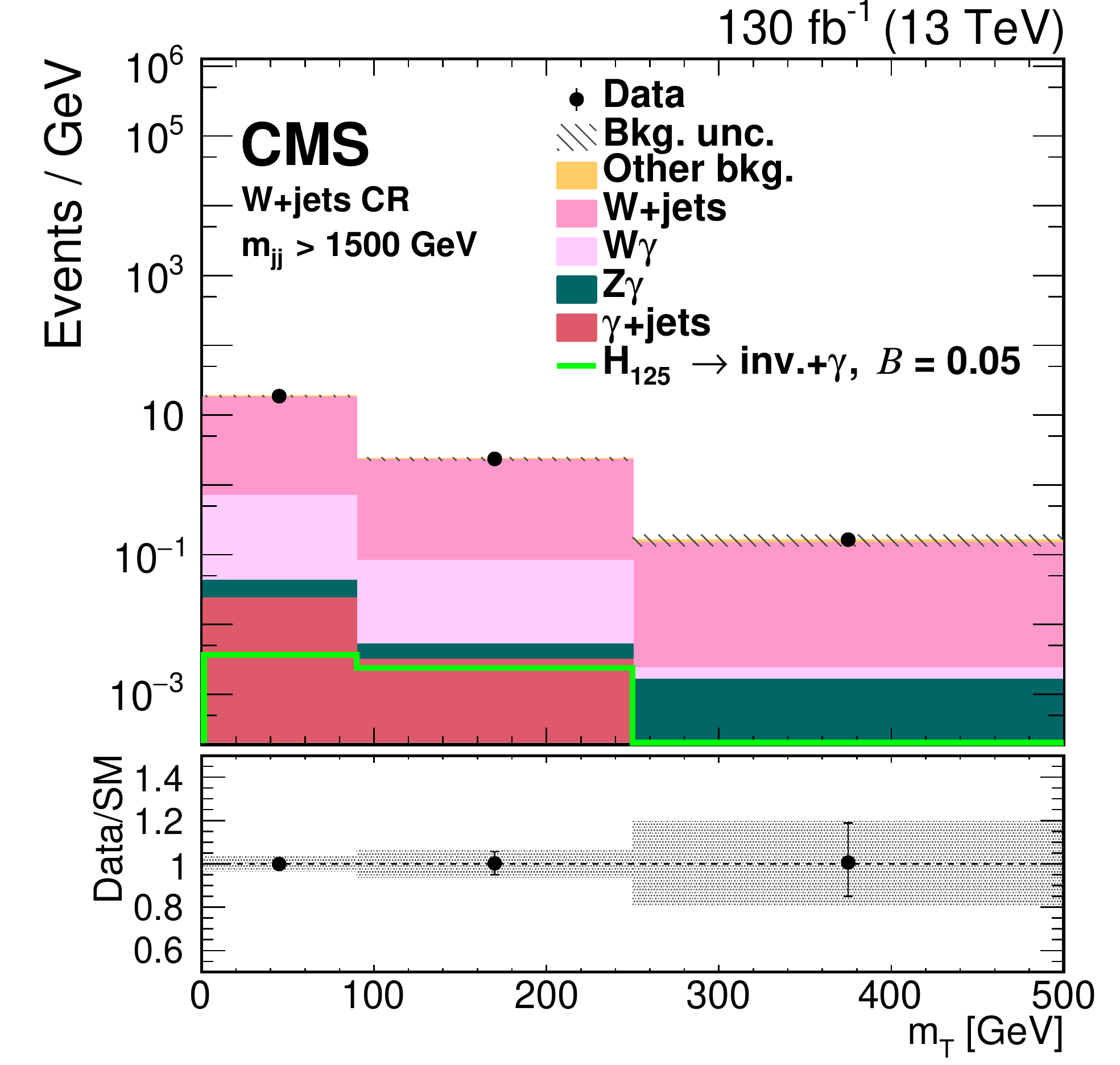}\\
\includegraphics[width=0.49\textwidth]{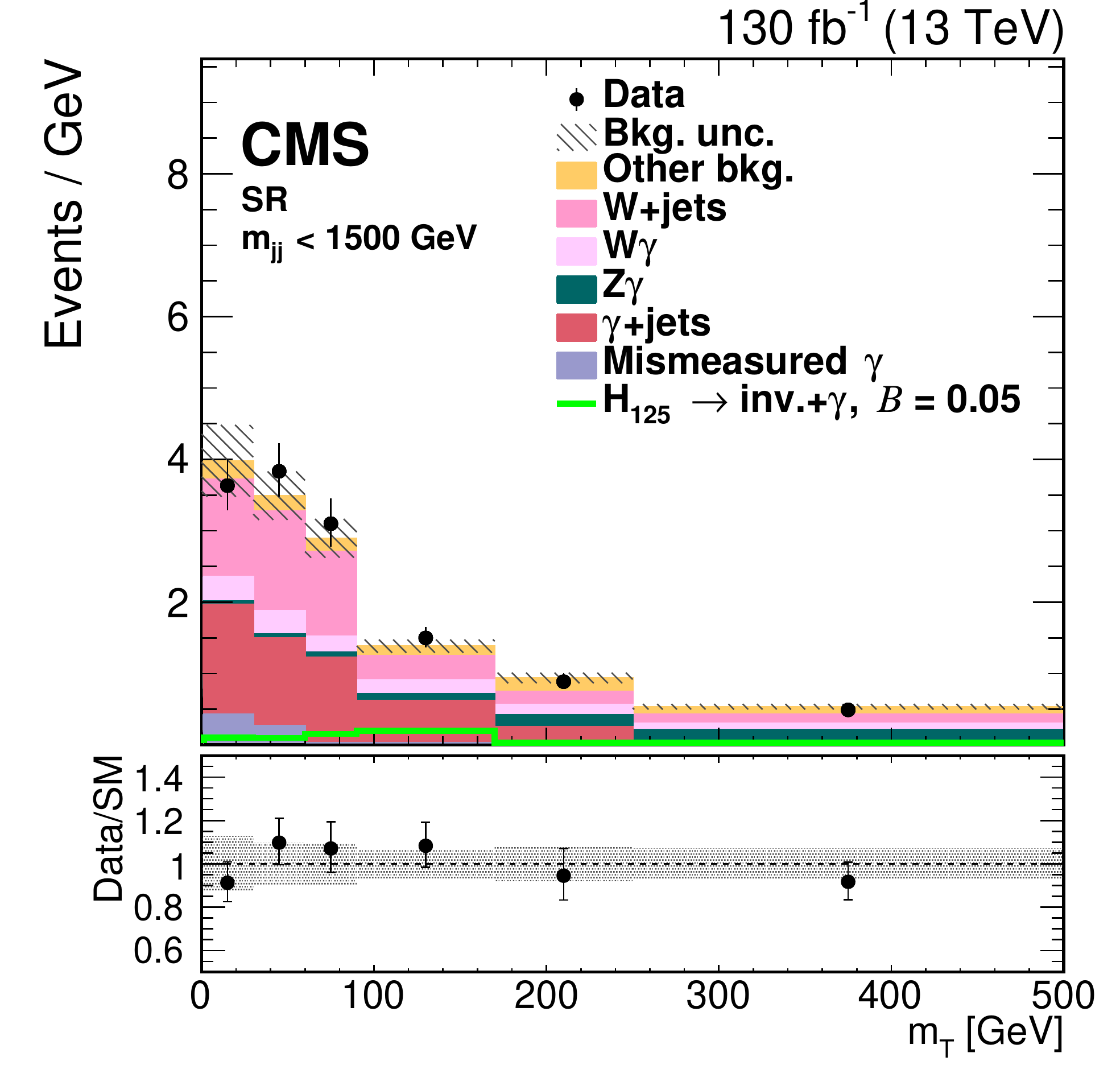}
\includegraphics[width=0.49\textwidth]{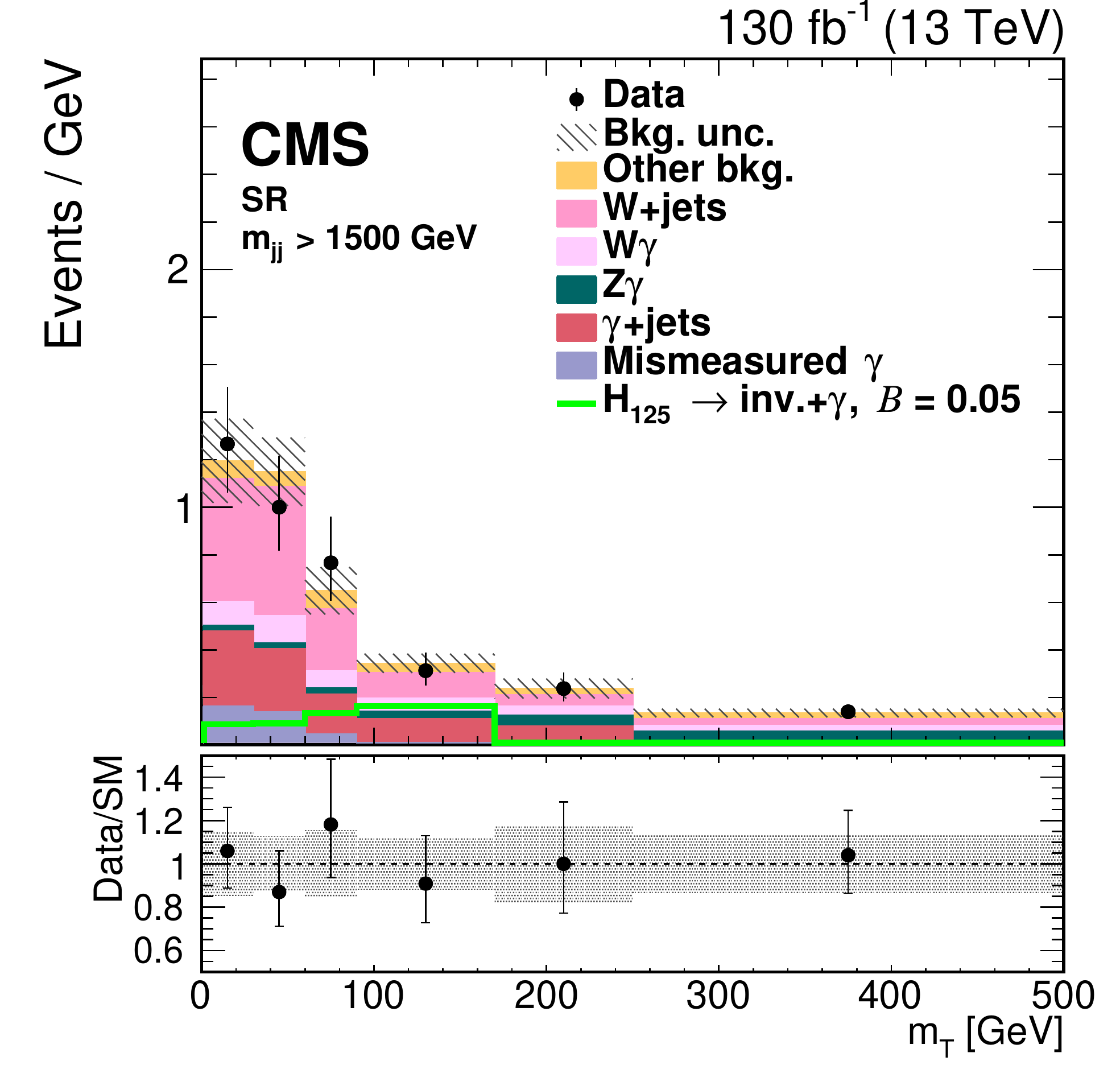}
 \caption{
The $\mT$ distributions from the simultaneous fit 
for events with $\mjj<1500\GeV$ in the $\wenj$ CRs (upper left), 
for events with $\mjj\geq1500\GeV$ in the $\wenj$ CRs (upper right), 
for events with $\mjj<1500\GeV$ in the SRs (lower left), 
and for events with $\mjj\geq1500\GeV$ in the SRs (lower right). The category other background
includes contributions from  $\zj$, nonprompt, top quark, $\PV\PV$, and $\PV\PV\PV$ processes.
Overflow events are included in the last bin. The shaded bands 
represent the combination of the statistical and systematic uncertainties in the predicted yields. 
The light green line, illustrating the possible contribution expected from inclusive 
SM Higgs boson production, assumes a branching fraction of 5\% for $\PH\to\text{inv.}+\gamma$ decays.
The lower panel in the figures shows a per-bin ratio of the data yield and the background expectation.
The shaded band corresponds to the combined systematic and statistical uncertainty in the background expectation.
\label{fig:vbfg_comb1}}
\end{figure}

The VBF signal reconstruction efficiency increases with $m_{\PH}$, with values of 0.2, 2.6, and 8.2\% 
for masses of 125, 300, and 1000\GeV, respectively. 
The inefficiency is driven by the $\ptmiss$ and photon $\pt$ requirements. 
The $\mjj$ distributions in the $\gj$, $\zmmg$, and $\wmng$ CRs are shown in 
Fig.~\ref{fig:vbfg_comb0}, while the  $\mT$ distributions in the $\wenj$ CRs and 
in the SRs are shown in Fig.~\ref{fig:vbfg_comb1}. 
The signal spectrum shows a Jacobian peak with an end-point at $\mT \sim m_{\PH}$, while the 
background processes have either a flat distribution or display an increase towards 
lower values of $\mT$.

\begin{table}[htbp]
  \centering

  \topcaption{Data, expected backgrounds, and estimated signal in the different regions. 
  The expected background yields are shown with their best-fit normalizations 
  from the simultaneous fit assuming background-only in the different regions. 
  The combination of the statistical and systematic uncertainties is shown. 
  The illustrative signal yield assumes a production cross section of $0.05\sigmaSM$. 
  All data-taking periods and trigger paths are combined together for each region.}
  \label{tab:postfityields}
\cmsTable{
  \begin{tabular} {lccccc}
\hline
 & SR & $\wenj$ CR & $\zmmg$ CR & $\wmng$ CR & $\gj$ CR \\
\hline
$\wj$            &   250   $\pm$   17   & 10500   $\pm$  100   & 	\NA	    &	     \NA	  &   180   $\pm$  37   \\
$\wlng$          &    98   $\pm$   11   &   240   $\pm$   36   & 	\NA	    &  190   $\pm$   18   &    76   $\pm$   8   \\
$\zg$            &    98   $\pm$   18   &     6.8 $\pm$    1.5 &    25  $\pm$   4   &    1.7 $\pm$    0.4 &    46   $\pm$   8   \\
$\gj$            &   230   $\pm$   22   &    12   $\pm$    4   & 	\NA	    &    9.5 $\pm$    2.3 &  1400   $\pm$  58   \\
Mism. $\gamma$   &    34   $\pm$   15   &	  \NA	       &  	\NA	    &	     \NA	  &	    \NA 	\\
$\zj$            &    41   $\pm$    6   &   100   $\pm$   10   & 	\NA	    &    6.3 $\pm$    0.6 &    26   $\pm$   3   \\
Nonprompt        &    20   $\pm$    4   &     1.1 $\pm$    0.2 &    1.2  $\pm$	0.2 &    4.4 $\pm$    0.9 &    62   $\pm$  13   \\
Top quark	 &    18   $\pm$    5   &    16   $\pm$    4   &    0.3  $\pm$  0.1 &	30   $\pm$    7   &    22   $\pm$   5   \\
$\PV\PV$	 &     6.9 $\pm$    1.0 &   200   $\pm$    9   &    0.3  $\pm$  0.3 &    4.4 $\pm$    0.9 &	5.7 $\pm$   0.5 \\
$\PV\PV\PV$	 &     3.1 $\pm$    0.5 &     7.6 $\pm$    1.0 &      \NA	    &    8.1 $\pm$    1.1 &     3.6 $\pm$   0.5 \\[\cmsTabSkip]

Total background &   800   $\pm$  25    & 11100   $\pm$  100   &    27    $\pm$  4   &  250   $\pm$   16  &  1800   $\pm$ 43   \\[\cmsTabSkip]

Data             &   801                & 11091                &    27               &  253               &     1830            \\[\cmsTabSkip]

$\Pq\Pq\PH_{125}(\gamma\gamma_\mathrm{D})$ &    50.5   $\pm$   7.4 &	 1.7 $\pm$   0.3  &   \NA	  &	     \NA  &	4.5 $\pm$   0.4 \\
$\Pg\Pg\PH_{125}(\gamma\gamma_\mathrm{D})$ &    30.6   $\pm$  14.3 &	 1.2 $\pm$   0.6  &   \NA	  &	     \NA  &	6.9 $\pm$   2.9 \\
  \hline
  \end{tabular}
}
\end{table}

\begin{figure}[hbt]
\centering
\includegraphics[width=0.65\textwidth]{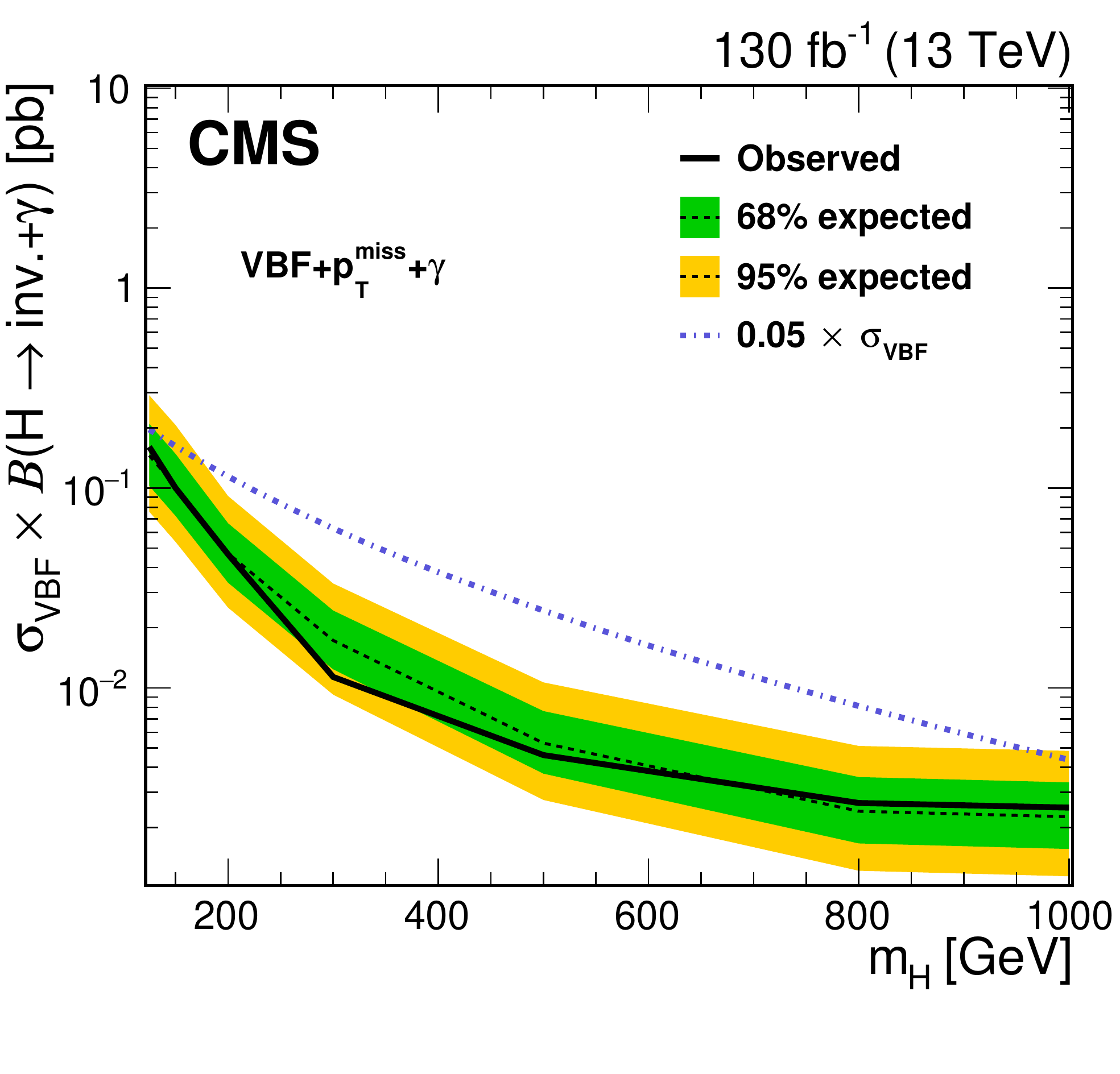}
 \caption{
   Expected and observed upper limits at 95\% $\CL$ on the product of 
   $\sigmavbf$ and $\brhinvg$ as a 
   function of $m_{\PH}$. 
   The dot-dashed line shows the predicted  signal corresponding to $0.05\sigmavbf$, assuming SM couplings.
   A linear interpolation is performed between the values obtained for the probed $m_{\PH}$ values. 
\label{fig:limits}}
\end{figure}

No significant excess of events above the expectation from the SM background is found. 
The upper limits at 95\% $\CL$ are calculated using a modified frequentist 
approach with the $\CLs$ criterion~\cite{Read1,Junk:1999kv} and an asymptotic 
method for the test statistic~\cite{LHC-HCG-Report,Cowan:2010js}. 
The statistical compatibility of the observed results, using the test based 
on a saturated $\chi^2$ model~\cite{Baker:1983tu}, with the expectation under 
the background-only hypothesis corresponds to a p-value of 0.25. The expected 
and observed cross section upper limits at 95\% $\CL$ on the product of the signal cross section $\sigmavbf$ for VBF production
and $\brhinvg$ as a function of $m_{\PH}$ are shown in Fig.~\ref{fig:limits}, and 
range from $\approx$160 to $\approx$2\unit{fb} as $m_{\PH}$ increases from 125 to 1000\GeV. 
For the years 2017 and 2018, the \ptmiss trigger path is the most sensitive one for 
signal models with $m_{\PH}\lesssim400\GeV$;  above this value, the single-photon trigger path dominates.
These limits also apply to other models where a scalar particle decays to a photon and light 
invisible particles.
For $m_{\PH}=125\GeV$, the result is interpreted as an upper limit on $\brhinvg$ 
assuming the production rate for an SM Higgs boson~\cite{deFlorian:2016spz}. 
In this case, the additional contribution from the $\Pg\Pg\PH$ production in the VBF 
category is considered, accounting for an increase in the signal yields of about 60\%, and mainly 
contributing to the region with $\mjj<1500\GeV$.
The observed (expected) 95\% $\CL$ upper limit at $m_{\PH}=125\GeV$ on 
$\brhinvg$ is 3.5 ($2.8^{+1.3}_{-0.8}$)\%.

The results of this analysis are combined with a complementary search for the same Higgs boson decay where the 
Higgs boson is produced in association with a $\PZ$ boson 
($\PZ\PH$)~\cite{Sirunyan:2019xst}. The combination is performed assuming the 
production rates for an SM-like 125\GeV Higgs boson~\cite{deFlorian:2016spz}.
For the combination, all the experimental uncertainties are treated as correlated between the two analyses, 
while all others are treated as uncorrelated. 
The observed and expected 95\% \CL limits at $m_{\PH}=125\GeV$ on $\brhinvg$ for the VBF category, $\PZ\PH$ category, 
and their combination are shown in Table~\ref{tab:comblimits}. The combined observed (expected) upper limit 
at 95\% \CL at $m_{\PH}=125\GeV$ on $\brhinvg$ is 2.9 (2.1)\%.

\begin{table}[htb]
  \centering
  \topcaption{Observed and expected 95\% \CL limits at $m_{\PH}=125\GeV$ on $\brhinvg$ 
  for the VBF category, $\PZ\PH$ category, and their combination.}
  \label{tab:comblimits}
  \begin{tabular} {cccccc}
\hline
\multicolumn{2}{c}{VBF} & \multicolumn{2}{c}{$\PZ\PH$} & \multicolumn{2}{c}{VBF+$\PZ\PH$} \\
Obs. (\%) & Exp. (\%) & Obs. (\%) & Exp. (\%) & Obs. (\%) & Exp. (\%) \\
\hline
3.5 & $2.8^{+1.3}_{-0.8}$ &  4.6 & $ 3.6^{+2.0}_{-1.2}$ & 2.9  & $2.1^{+1.0}_{-0.7}$ \\
  \hline
  \end{tabular}
\end{table}

\section{Summary}
\label{sec:summary}

A search has been presented for a Higgs boson that is produced via vector boson fusion (VBF) and that 
decays to an undetected particle and a photon. This is the first analysis for such decays in the VBF channel. 
The search has been performed by the CMS Collaboration using a data set corresponding 
to an integrated luminosity of 130\fbinv recorded at a center-of-mass energy of 13\TeV in 2016-2018. 
No significant excess of events above the expectation from the standard model 
background is found. 
The results are used to place limits on the product of the signal cross section $\sigmavbf$ for VBF production 
and the branching fraction for such decays of the Higgs boson, in the context of a 
theoretical model where the undetected particle is a massless dark photon. 
Allowing for deviations from standard model VBF production, the upper limit on the product of 
$\sigmavbf$ and $\brhinvg$ ranges from $\approx$160 to $\approx$2\unit{fb}, 
for $m_{\PH}$ from 125\GeV to 1000\GeV. 
The observed (expected) upper limit at 95\% confidence level 
at $m_{\PH}=125\GeV$ assuming standard model production rates on $\brhinvg$ is 3.5 (2.8)\% for this channel. 
Combining with an existing analysis targeting associated $\PZ$ boson 
production, and assuming the standard model rates, the observed (expected) upper 
limit at 95\% confidence level at $m_{\PH}=125\GeV$ on $\brhinvg$ is 2.9 (2.1)\%. 

\begin{acknowledgments}
  We congratulate our colleagues in the CERN accelerator departments for the excellent performance of the LHC and thank the technical and administrative staffs at CERN and at other CMS institutes for their contributions to the success of the CMS effort. In addition, we gratefully acknowledge the computing centers and personnel of the Worldwide LHC Computing Grid for delivering so effectively the computing infrastructure essential to our analyses. Finally, we acknowledge the enduring support for the construction and operation of the LHC and the CMS detector provided by the following funding agencies: BMBWF and FWF (Austria); FNRS and FWO (Belgium); CNPq, CAPES, FAPERJ, FAPERGS, and FAPESP (Brazil); MES (Bulgaria); CERN; CAS, MoST, and NSFC (China); COLCIENCIAS (Colombia); MSES and CSF (Croatia); RIF (Cyprus); SENESCYT (Ecuador); MoER, ERC IUT, PUT and ERDF (Estonia); Academy of Finland, MEC, and HIP (Finland); CEA and CNRS/IN2P3 (France); BMBF, DFG, and HGF (Germany); GSRT (Greece); NKFIA (Hungary); DAE and DST (India); IPM (Iran); SFI (Ireland); INFN (Italy); MSIP and NRF (Republic of Korea); MES (Latvia); LAS (Lithuania); MOE and UM (Malaysia); BUAP, CINVESTAV, CONACYT, LNS, SEP, and UASLP-FAI (Mexico); MOS (Montenegro); MBIE (New Zealand); PAEC (Pakistan); MSHE and NSC (Poland); FCT (Portugal); JINR (Dubna); MON, RosAtom, RAS, RFBR, and NRC KI (Russia); MESTD (Serbia); SEIDI, CPAN, PCTI, and FEDER (Spain); MOSTR (Sri Lanka); Swiss Funding Agencies (Switzerland); MST (Taipei); ThEPCenter, IPST, STAR, and NSTDA (Thailand); TUBITAK and TAEK (Turkey); NASU (Ukraine); STFC (United Kingdom); DOE and NSF (USA).
  
  \hyphenation{Rachada-pisek} Individuals have received support from the Marie-Curie program and the European Research Council and Horizon 2020 Grant, contract Nos.\ 675440, 752730, and 765710 (European Union); the Leventis Foundation; the A.P.\ Sloan Foundation; the Alexander von Humboldt Foundation; the Belgian Federal Science Policy Office; the Fonds pour la Formation \`a la Recherche dans l'Industrie et dans l'Agriculture (FRIA-Belgium); the Agentschap voor Innovatie door Wetenschap en Technologie (IWT-Belgium); the F.R.S.-FNRS and FWO (Belgium) under the ``Excellence of Science -- EOS" -- be.h project n.\ 30820817; the Beijing Municipal Science \& Technology Commission, No. Z191100007219010; the Ministry of Education, Youth and Sports (MEYS) of the Czech Republic; the Deutsche Forschungsgemeinschaft (DFG) under Germany's Excellence Strategy -- EXC 2121 ``Quantum Universe" -- 390833306; the Lend\"ulet (``Momentum") Program and the J\'anos Bolyai Research Scholarship of the Hungarian Academy of Sciences, the New National Excellence Program \'UNKP, the NKFIA research grants 123842, 123959, 124845, 124850, 125105, 128713, 128786, and 129058 (Hungary); the Council of Science and Industrial Research, India; the HOMING PLUS program of the Foundation for Polish Science, cofinanced from European Union, Regional Development Fund, the Mobility Plus program of the Ministry of Science and Higher Education, the National Science Center (Poland), contracts Harmonia 2014/14/M/ST2/00428, Opus 2014/13/B/ST2/02543, 2014/15/B/ST2/03998, and 2015/19/B/ST2/02861, Sonata-bis 2012/07/E/ST2/01406; the National Priorities Research Program by Qatar National Research Fund; the Ministry of Science and Higher Education, project no. 02.a03.21.0005 (Russia); the Tomsk Polytechnic University Competitiveness Enhancement Program; the Programa Estatal de Fomento de la Investigaci{\'o}n Cient{\'i}fica y T{\'e}cnica de Excelencia Mar\'{\i}a de Maeztu, grant MDM-2015-0509 and the Programa Severo Ochoa del Principado de Asturias; the Thalis and Aristeia programs cofinanced by EU-ESF and the Greek NSRF; the Rachadapisek Sompot Fund for Postdoctoral Fellowship, Chulalongkorn University and the Chulalongkorn Academic into Its 2nd Century Project Advancement Project (Thailand); the Kavli Foundation; the Nvidia Corporation; the SuperMicro Corporation; the Welch Foundation, contract C-1845; and the Weston Havens Foundation (USA).
\end{acknowledgments}

\bibliography{auto_generated}
\cleardoublepage \appendix\section{The CMS Collaboration \label{app:collab}}\begin{sloppypar}\hyphenpenalty=5000\widowpenalty=500\clubpenalty=5000\vskip\cmsinstskip
\textbf{Yerevan Physics Institute, Yerevan, Armenia}\\*[0pt]
A.M.~Sirunyan$^{\textrm{\dag}}$, A.~Tumasyan
\vskip\cmsinstskip
\textbf{Institut f\"{u}r Hochenergiephysik, Wien, Austria}\\*[0pt]
W.~Adam, T.~Bergauer, M.~Dragicevic, J.~Er\"{o}, A.~Escalante~Del~Valle, R.~Fr\"{u}hwirth\cmsAuthorMark{1}, M.~Jeitler\cmsAuthorMark{1}, N.~Krammer, L.~Lechner, D.~Liko, I.~Mikulec, F.M.~Pitters, N.~Rad, J.~Schieck\cmsAuthorMark{1}, R.~Sch\"{o}fbeck, M.~Spanring, S.~Templ, W.~Waltenberger, C.-E.~Wulz\cmsAuthorMark{1}, M.~Zarucki
% ...
\vskip\cmsinstskip
\dag: Deceased\\
1:  Also at Vienna University of Technology, Vienna, Austria\\
% ...
\end{sloppypar}
\end{document}